\documentclass[english, aps, prl, preprint]{revtex4-2}
\usepackage{units}
\usepackage{color}
\usepackage{amstext}
\usepackage{amssymb}
\usepackage{amsmath}
\usepackage{textcomp}
\usepackage{graphicx}
\usepackage{epstopdf}
\usepackage[T2A]{fontenc}
\usepackage{upgreek}
\usepackage{hyperref}
\usepackage{orcidlink}
\usepackage{gensymb}

\begin{document}
\title{Transport Detection of Whirlpools in GaAs Electron Liquid}

\author{Dmitry~A.~Egorov$^{1,2}$\,\orcidlink{0000-0001-5668-7284}}
\email[Contact author:~]{d.egorov@g.nsu.ru}

\author{Dmitriy~A.~Pokhabov$^{1,2}$\,\orcidlink{0000-0002-8747-0261}}

\author{Evgeny~Yu.~Zhdanov$^{1,2}$\,\orcidlink{0000-0002-7173-6213}}

\author{Andrey~A.~Shevyrin$^{1}$\,\orcidlink{0000-0003-0632-2636}}

\author{Askhat~K.~Bakarov$^{1,2}$\,\orcidlink{0000-0002-0572-9648}}

\author{Arthur~G.~Pogosov$^{1,2}$\,\orcidlink{0000-0001-5310-4231}}

\affiliation{$^{1}$Rzhanov Institute of Semiconductor Physics SB RAS, 13 Lavrentiev Ave., Novosibirsk, 630090, Russia}
\affiliation{$^{2}$Department of Physics, Novosibirsk State University, 2 Pirogov Str., Novosibirsk, 630090, Russia}

\date{\today}

\begin{abstract}
We report the formation of large-scale steady-state whirlpools in a GaAs-based two-dimensional electron liquid and demonstrate them by straightforward transport measurements. A whirlpool forming inside a circular cavity adjoining a wide conducting channel appears as a negative four-terminal resistance over a broad range of temperatures and cavity sizes. The effect scales with the Gurzhi length, in quantitative accord with the hydrodynamic analogy. Obtained results firmly establish this analogy and probe the limits of its applicability.
\end{abstract}

\maketitle

The hydrodynamic approach to electron transport \cite{gurzhi1968}, in which electrons in pure conductors are treated under appropriate conditions as a viscous incompressible fluid, owes its fruitfulness to the transparent analogy between momentum transfer processes in an electron system and those in classical fluids. The key parameter governing hydrodynamic effects is the electron viscosity \cite{alekseev2020, narozhny2022, Fritz2024}, which is controlled by electron-electron (e-e) interactions. Viscous effects emerge at moderately high temperatures, where the e-e scattering rate is sufficiently high, yet phonon scattering has not yet disrupted the collective nature of the flow. The development of this hydrodynamic analogy has stimulated predictions of a broad range of qualitatively new phenomena, subsequently confirmed experimentally in sufficiently clean electron systems such as the two-dimensional electron gas (2DEG) in graphene and GaAs. Among the most prominent manifestations of electron hydrodynamics are the Gurzhi effect \cite{gurzhi1962, dejong1995}, superballistic transport \cite{guo2017, krishnakumar2017, ginzburg2021, Sarypov2025, Estradalvarez2025}, Poiseuille \cite{Sulpizio2019, Ku2020, Vool2021} and slip electron \cite{Sarypov2025Slip} velocity profiles, turbulent phenomena \cite{Mendoza2011, Sarypov2026}, and Hall viscosity \cite{Scaffidi2017, Pellegrino2017, gusev2018, Berdyugin2019, BenShachar2025}.

Vortices are among the most striking manifestations of electron hydrodynamics \cite{Pellegrino2016, Danz2020}. Indirect evidence for their existence, most notably negative four-terminal resistance measured in the vicinity geometry, has been reported in graphene \cite{bandurin2016, bandurin2018} and GaAs \cite{levin2018, Gupta2021PRL}. However, negative resistance alone does not constitute unambiguous evidence of the hydrodynamic regime, as it can also arise in the ballistic limit. Direct experimental observation of hydrodynamic vortices has therefore long remained elusive. Only recently have electron whirlpools been visualized using specialized probe techniques and in a limited set of materials: specifically, in $\textrm{WTe}_2$ single crystals employing scanning SQUID microscopy \cite{AharonSteinberg2022} and in graphene at room temperature using scanning NV magnetometry \cite{Palm2024}. As a result, electron vortices remain poorly characterized experimentally, owing to the complexity of the required instrumentation and certain sample size constraints inherent to the materials employed.

\begin{figure*}
\centering
\includegraphics[width=1\linewidth]{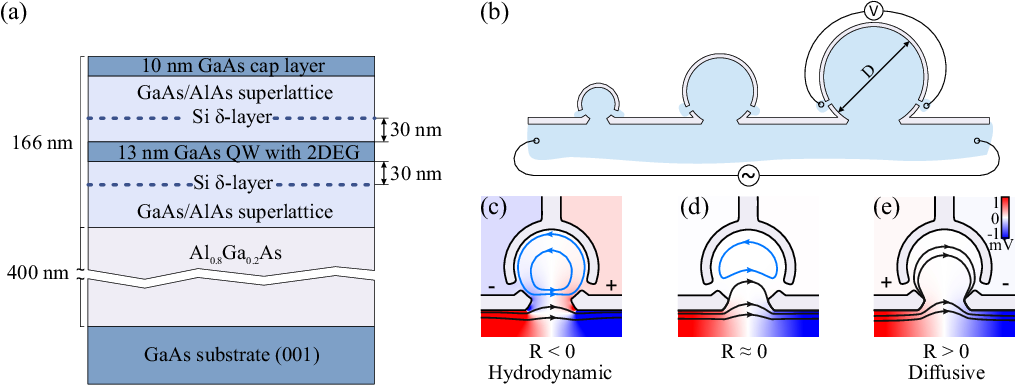}
\caption{\label{fig:Fig_1} 
(a) Schematic of the heterostructure with a 2DEG in a GaAs quantum well. (b) Schematic of the studied circular cavities with diameters $D=8$, $16$, and $24$~$\upmu$m adjoining a wide channel. Current streamlines of the electron liquid in a circular cavity in the (c) hydrodynamic, (d) transitional, and (e) diffusive transport regimes, and the corresponding calculated potential maps. }
\end{figure*}

In this work, we implement an alternative, purely transport-based approach to detecting steady-state electron whirlpools in a GaAs-based two-dimensional electron gas. In our structures, the Gurzhi length --- the characteristic length scale of hydrodynamic effects --- can reach $10$~$\upmu$m, substantially exceeding its values in graphene, making it feasible to observe vortices in large circular cavities equipped with potentiometric contacts. We show that when current is passed through the main channel, a robust negative resistance develops between the contacts attached to the cavity in the linear response regime, which is naturally interpreted as a signature of steady-state circulatory flow of the electron liquid. By studying the evolution of this response with temperature and cavity diameter, we can disentangle the ballistic and hydrodynamic contributions and find that the sign reversal of the resistance occurs at a universal ratio between the cavity diameter $D$ and the Gurzhi length $D_{\upnu}$, which sets the characteristic scale of hydrodynamic effects \cite{Torre2015}. 

\textit{Samples for studying steady-state vortices ---} The experimental samples are based on a high-mobility 2DEG hosted in GaAs/AlGaAs heterostructures (Fig. \ref{fig:Fig_1} (a)) \cite{friedland1996}. At $4.2$~K, the electron density and mobility are $n = 7 \times 10^{11}$~cm$^{-2}$ and $\upmu \approx 6 \times 10^5$~cm$^2$/(V $\cdot$ s), respectively. Circular cavities adjoining a wide ($50$~$\upmu$m) 2DEG channel were defined by lithography (Fig. \ref{fig:Fig_1} (b)). The cavity diameters $D$ are $8$, $16$, and $24$~$\upmu$m. The lithographic width of the cavity opening is equal to its radius. Each cavity is equipped with two microcontacts to probe the local potential. The lithographic width of the contacts is 0.8~$\upmu$m for all cavities, regardless of their diameter. The contacts are placed symmetrically on either side of the cavity opening, subtending a central angle of $120\degree$.

The potential difference $V$ between the potentiometric contacts in circular cavities of various diameters was measured as a function of the DC current passed along the main channel. Measurements were performed at currents up to $1$~mA and over a temperature range from $3.5$ to $90$~K, enabling a systematic study of the evolution of the response across a wide range of the key physical parameters governing the observed effect: the electron-electron scattering length $l_{\textrm{ee}}$, the momentum relaxation length $l_{\textrm{mr}}$, the viscosity $\nu$, and the Gurzhi length $D_{\upnu}$.

\begin{figure*}
\centering
\includegraphics[width=1\linewidth]{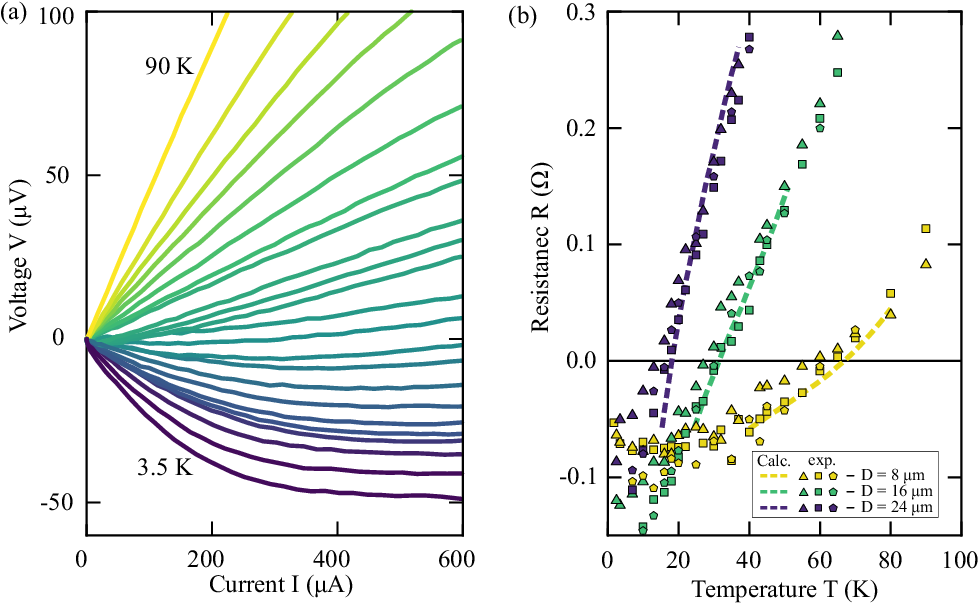}
\caption{\label{fig:Fig_2} 
(a) Potential difference $V$ in a circular cavity of diameter $D=16$~$\upmu$m as a function of the DC current $I$ passed through the channel, measured at various temperatures. (b) Four-terminal resistance as a function of temperature for circular cavities of different diameters: $8$~$\upmu$m (yellow, light in grayscale), $16$~$\upmu$m (green, gray in grayscale), and $24$~$\upmu$m (purple, dark in grayscale). Different symbols of the same color correspond to experimentally measured slopes obtained from different devices of identical geometry. Dashed lines show the results of numerical simulations. }
\end{figure*}

\textit{Potential measurements inside the cavity ---} Fig. \ref{fig:Fig_2} (a) shows a representative set of potential difference $V$ versus current $I$ curves measured at various temperatures for a cavity of diameter $D = 16$~$\upmu$m. Note, at small currents the $V$-$I$ relationship is linear. At low temperatures, a negative voltage $V<0$ is observed, corresponding to a negative resistance $R=V/I$. As shown below, this is a signature of steady-state vortex formation inside the cavity. As the temperature increases, the negative response is gradually replaced by a positive one. At elevated temperatures, the $V$-$I$ dependence becomes nearly linear across the entire range of currents investigated. Qualitatively similar behavior is observed for all studied samples with different cavity diameters.

From the measured set of curves, the temperature dependence of the four-terminal resistance $R$ was extracted in the low-current linear response regime. The obtained temperature dependences of $R$ are presented in Fig. \ref{fig:Fig_2} (b). Data points of different shapes correspond to measurements from three independent samples, each containing cavities of all three studied diameters: $D=8$, $16$, and $24$~$\upmu$m. A clear evolution of the resistance is observed, from a negative response at low temperatures to conventional positive resistance upon heating.

All studied samples exhibit a robust region of negative resistance, with the temperature extent of this region depending strongly on the cavity diameter $D$. The resistance in samples with $D=16$~$\upmu$m and $D=24$~$\upmu$m increases nearly linearly with temperature across the entire range (Fig. \ref{fig:Fig_2} (b)). In contrast, for samples with $D=8$~$\upmu$m the resistance increases only at $T > 30$~K, whereas at $T < 30$~K the response is nearly temperature-independent.

\textit{Discussion ---} The sign of the response is determined by the direction of current flow inside the cavity relative to that in the main channel and depends on the electron transport regime. The positive resistance observed at high temperatures indicates that these directions coincide, consistent with the conventional Ohmic regime, characterized by frequent electron scattering off phonons and impurities and, consequently, momentum relaxation lengths $l_{\textrm{mr}}$ that are short compared to the characteristic sample dimensions. Negative resistance, on the other hand, is a signature of reversed current flow inside the cavity and can be attributed either to ballistic electron traversal or to the formation of hydrodynamic vortices within the cavity.

Ballistic traversal occurs at the lowest temperatures, when the momentum relaxation length $l_\textrm{mr}$ exceeds the characteristic cavity dimensions. For the $D=8$~$\upmu$m sample, this condition is met below $T \sim 30$~K \cite{giuliani1982, chaplik1971, Egorov2024}. Under such conditions, electrons traverse a cavity without momentum-relaxing scattering, which naturally explains the resistance saturation observed in this regime. For larger cavities whose diameters exceed the low-temperature $l_\textrm{mr}$ ballistic traversal does not occur in the whole temperature range (Fig. \ref{fig:Fig_2} (b)). Accordingly, the resistance behavior in the $D=8$~$\upmu$m sample at $T \gtrsim 30$~K is governed by the interplay between hydrodynamic vortex flow and diffusive transport, as is the case for the larger samples across the entire temperature range.

\begin{figure}
\centering
\includegraphics[width=0.5\linewidth]{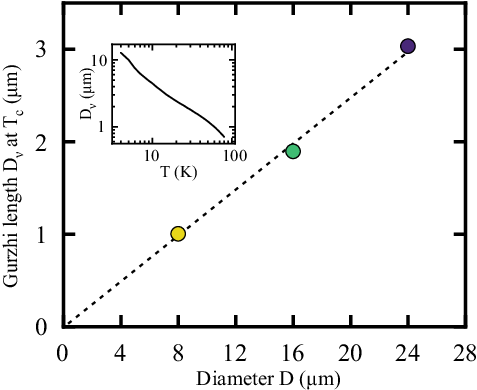}
\caption{\label{fig:Fig_3} 
Gurzhi length $D_\upnu$ at the transition temperature $T_\textrm{c}$ as a function of the circular cavity diameter. The left inset shows the temperature dependence of $D_\upnu$ in our samples. }
\end{figure}

Let us consider the transition points between negative and positive resistance (Fig. \ref{fig:Fig_2} (b)). The sign reversal of the resistance marks the disappearance of vorticity inside the cavity. For samples with diameters $D=8$, $16$, and $24$~$\upmu$m, the transition occurs at critical temperatures $T_\textrm{c} \approx 60$, $30$, and $16$~K, respectively. Note that the critical temperature $T_c$ scales non-linearly with the cavity diameter $D$. The key parameter governing the transition proves to be the Gurzhi length $D_{\upnu}$. Indeed, as seen from Fig. \ref{fig:Fig_3}, the observed effect scales clearly with the Gurzhi length $D_{\upnu}$ evaluated at the critical temperature $T_c$, obeying the simple relation $D/D_{\upnu} \approx 8.0$. This universal scaling with the Gurzhi length further corroborates the hydrodynamic nature of the observed transitions. A similar scaling, with a slightly different value of the ratio $D/D_\upnu$, has previously been reported in other material systems \cite{Palm2024, AharonSteinberg2022}.

It is worth noting that the Gurzhi length
\begin{equation}\label{D_nu}
D_{\upnu} = \sqrt{\nu\tau_\textrm{mr}}=\sqrt{\nu \frac{l_\textrm{mr}}{v_{\textrm{F}}}}, 
\end{equation}
used in the analysis of the experimental data was obtained from the electron viscosity $\nu$ and the momentum relaxation time $\tau_\textrm{mr}$, both experimentally determined in the heterostructure studied in this work, from the superballistic conductance of microconstrictions \cite{Sarypov2025} and the longitudinal magnetoresistance \cite{Zhdanov2026}, respectively. The temperature dependence of the Gurzhi length is shown in the left inset of Fig. \ref{fig:Fig_3}.

It is interesting to note that an alternative expression for the Gurzhi length, $\frac{1}{2}\sqrt{l_{\textrm{ee}}l_{\textrm{mr}}}$, can be obtained from \eqref{D_nu} using the widely adopted relation $\nu = \frac{1}{4}v_{\text{F}}l_{\text{ee}}$, with $l_{\textrm{ee}}$ either calculated according to theory \cite{chaplik1971, giuliani1982} or measured in the heterostructure under study by transverse magnetic focusing \cite{Egorov2024, Egorov2025}. However, this alternative expression does not yield the clear scaling observed when using \eqref{D_nu}. This points to a breakdown of the commonly used relation $\nu = \frac{1}{4}v_{\text{F}}l_{\text{ee}}$ and highlights the necessity of employing directly measured values of the viscosity rather than the electron-electron scattering length. Similar conclusions have been drawn in recent alternative experimental studies \cite{keser2021, Sarypov2025, https://doi.org/10.48550/arxiv.2602.16847}.

To illustrate the crossover between the transport regimes discussed above, numerical simulations were carried out within the hydrodynamic model. Fig. \ref{fig:Fig_1} (c-e) shows the calculated potential distribution maps and current streamlines for a sample with $D=8$~$\upmu$m at various Gurzhi lengths. At a large Gurzhi length (Fig. \ref{fig:Fig_1} (c), $D_{\upnu} > D/8$), the simulation reveals a hydrodynamic vortex inside the cavity, with a potential distribution corresponding to a negative response. As the Gurzhi length decreases (Fig. \ref{fig:Fig_1} (d), $D_{\upnu} \gtrsim D/8$), the current from the channel expels the vortex and the potential difference approaches zero. At small Gurzhi length (Fig. \ref{fig:Fig_1} (e), $D_{\upnu} < D/8$), no vortex forms and transport becomes conventional and ohmic. The calculated resistances for each sample are overlaid on Fig. \ref{fig:Fig_2} (b) and are in good agreement with the experimental data, further supporting the interpretation of the negative response as a signature of a steady-state hydrodynamic vortex. Note that the simulations were performed in the slip boundary condition limit, which provides the best agreement with the experimental data and is consistent with the perfect slip electron flow previously reported in GaAs heterostructures \cite{Sarypov2025Slip}.

\begin{figure}
\centering
\includegraphics[width=0.5\linewidth]{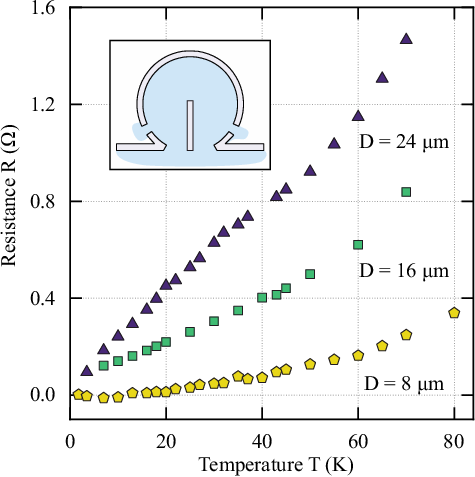}
\caption{\label{fig:Fig_4}
Four-terminal resistance as a function of temperature for circular cavities of different diameters with an etched strip (see inset). }
\end{figure}

As an additional consistency check, we also performed a reference experiment on samples with circular cavities of various diameters containing a narrow etched strip (see inset of Fig. \ref{fig:Fig_4}). This geometry inhibits the formation of a hydrodynamic vortex. Indeed, as shown in Fig. \ref{fig:Fig_4}, the resistance is positive in all studied samples across the entire temperature range, confirming the absence of vortices. At the same time, the observed dependences qualitatively reproduce the key features of the main experiment: almost linear temperature trends of the resistance, and, in the smallest sample, a low-temperature saturation attributable to the ballistic contribution.

\textit{Conclusion ---} In summary, we have experimentally demonstrated the formation of steady-state large-scale hydrodynamic vortices in an electron liquid using conventional transport measurements. Employing a simple lithographic geometry --- a circular cavity adjoining a wide channel --- and measuring the potential difference near the cavity opening, we observe a clear signature of vortex flow: a negative resistance.

A systematic study of samples with different cavity diameters over a wide temperature range allowed us to trace the evolution of the transport regime from ballistic, observed at $T<30$~K in small cavities, through hydrodynamic (negative response) to diffusive (positive response). One of the key results providing unambiguous evidence for the hydrodynamic interpretation is the universal scaling of the effect with the Gurzhi length at the critical transition temperature (at which the vortex disappears): the relation $D/D_{\upnu} \approx 8.0$ holds for all studied diameters. Notably, a consistent description of the observed effects is achieved only when the Gurzhi length is defined through the experimentally determined viscosity $\nu$ and momentum relaxation time $\tau_\textrm{mr}$. Attempts to employ the widely used relation $\nu = \frac{1}{4}v_{\text{F}}l_{\text{ee}}$ break the observed scaling, pointing to the need to revisit the relation. Numerical simulations within the hydrodynamic model are in quantitative agreement with the experimental data, further justifying the hydrodynamic interpretation.

The proposed transport approach thus offers a direct and relatively simple route to detecting whirlpools in an electron liquid. The observation of clear universal scaling across a wide range of temperatures and cavity sizes not only confirms the hydrodynamic nature of the negative response, but also demonstrates that the electron liquid in high-mobility GaAs heterostructures constitutes an ideal model system for the study of viscous effects. The ability to control the vortex state by varying temperature and geometry, combined with direct resistive detection of its signature, lays the groundwork for systematic studies of electron hydrodynamics and opens prospects for the development of novel device concepts operating on the principles of vortex charge flow control.

\medskip
\textit{Acknowledgements ---} The work is supported by the Russian Science Foundation (grant No 22-12-00343-$\Pi$ -- experimental study).

\section*{References}

\bibliography{Manuscript}

\end{document}